%% file: heraff.tex
\documentstyle[12pt,epsfig]{article}
\input{defi}
\begin{document}
%
%
%
%
\begin{flushright}
TAUP 2464 - 97\\
DESY 97 - 212 \\
Octorber 1997\\
\end{flushright}

~

~
\begin{center}

{\Large\bf{SCALING VIOLATION AND}}\\[1ex]
{\Large \bf {  SHADOWING CORRECTIONS AT HERA.}}\\[6ex]
{\large \bf { A. L.
Ayala  Filho${}^{a)\,b)}$${}^*$\footnotetext{ ${}^*$ E-mail:
{\tt ayala@if.ufrgs.br}},
 M. B. Gay  Ducati ${}^{a)}$$^{**}$\footnotetext{${}^{**}$
E-mail:{\tt gay@if.ufrgs.br}}
  and 
 E. M. Levin ${}^{c)\,d)\,\dagger}$
\footnotetext{$^{\dagger}$ E-mail: {\tt leving@post.tau.ac.il}} 
}} \\[6ex]
{\it ${}^{a)}$Instituto de F\'{\i}sica, Univ. Federal do Rio Grande do Sul}
{\it Caixa Postal 15051, 91501-970 Porto Alegre, RS, BRAZIL}\\
{\it ${}^{b)}$Instituto de F\'{\i}sica e Matem\'atica, Univ. 
Federal de Pelotas}
{\it Campus Universit\'ario, Caixa Postal 354, 96010-900, Pelotas, RS,
BRAZIL}\\
{\it ${}^{c)}$ HEP Department, School of Physics and Astronomy}\\
{\it Raymond and Beverly  Sackler Faculty of Exact Science}\\
{\it Tel Aviv University, Tel Aviv 69978, ISRAEL}\\

{\it$ {}^{d)}$ DESY Theory, Notkestr. 85, 22607, Hamburg, GERMANY }\\[3ex]
\end{center}
{\large \bf Abstract:} We study the value of  shadowing corrections (SC) 
 in  HERA kinematic region in Glauber - Mueller approach. Since the
Glauber - Mueller approach was proven in perturbative QCD in the double
logarithmic approximation (DLA), we develop the DLA approach for deep
inelastic structure function which takes into account the SC.
Our estimates show  small SC for $F_2$ in HERA kinematic region while 
 they turn out to be sizable for the gluon structure function. We compare
our estimates with those for gluon distribution in leading order (LO) and
next to leading order (NLO) in the DGLAP evolution equations.
\vspace{3cm}

In this letter, we investigate the role of the shadowing corrections (SC)
 on the value of the deep inelastic structure functions and on the
scaling 
violation mechanism in HERA kinematic region. Our estimates of the SC 
are done using the Glauber - Mueller formula. This formula was
proven by Mueller \cite{MU90} in the DLA of pQCD  and was studied in
details in Refs.\cite{frst}\cite{AGLA}\cite{AGLP}.   
The structure function $F_2(x,Q^2)$   reads as:
\bea
F_2(x, Q^2)= \frac{N_c}{6 \pi^3} \sum^{N_f} Z^2_f 
\int_{\frac{1}{Q^2}}^{\infty}\,\frac{d r_{\perp}^2}{r_{\perp}^4}\,
\int\,  d^2 b_{\perp}  \{1 - e^{- \frac{1}{2} \Omega } \}\, ,
\label{GMF}
\eea
where $x$ is the Bjorken scaling variable and $Q^2$ is the photon
virtuality.
$N_c$ is the colour number and $Z_f$ is the charge fraction of each quark,
$N_f$ is the number of flavours taken into account in the 
quark- nucleon scattering and  $b_{\perp}$ is the impact parameter for the
scattering of the quark - antiquark pair with transverse splitting
$r_{\perp}$ with the nucleon target.

The opacity function $\Omega$ is, generally speaking, an arbitrary real
function and it  has a simple physical interpretation: $e^{- \Omega}$
is the probability that the quark-antiquark  pair do not suffer an
inelastic scattering. This function  should be determined in
QCD. For the DLA it was shown ( see Refs.\cite{MU90}
\cite{LR87}) that the impact parameter dependence can be
factorized out and
 $\Omega$ can be written
\bea
\Omega(x, Q^2, b_{\perp})=\frac{4 \pi^2 \as}{3 Q^2} xG(x,Q^2) S(b_{\perp})
\label{ome}
\eea
where $S(b_{\perp})$ is the profile function for a nucleon with
radius $R$, and is
taken as an exponential function, 
$S(b_{\perp})=\frac{1}{\pi R^2}exp( -b^2_{\perp}/ R^2 )$, in our
calculations.

It should be stressed that \eq{GMF} takes into account the unitarity
constraint and with $\Omega$ defined in \eq{ome} gives the DGLAP evolution
equation \cite{DGLAP} in the kinematic region where $\Omega\,\,\ll\,\,1$.

In order to investigate the $x$ and $Q^2$ evolution of the gluon
distribution in this approach, we will consider first our Born term
with the gluon distribution taken in the DLA limit. Since the Born
term is equivalent to the DGLAP expression for $F_2$ in the DLA limit, 
we will take $xG^{DLA}$ as in Ref.\cite{bafo} and \cite{ina}. Thus,
the gluon distribution reads
\bea
xG^{DLA} (x, Q^2) = G_0 I_0 (y) \, ,
\label{xgdla}
\eea  
where the variables are $y=2 \gamma \sqrt{ ln (1/x) ln (t/t_0)}$ and 
$t= ln (Q^2/\Lambda^2)$.  The QCD constants are 
$\gamma =\sqrt{12/\beta_0}$,
$\beta_0 = 11 -  \frac{2 N_f}{3}$ and $\Lambda= 0.232 \, GeV^2$ .
The constant $G_0$ plays the role of the flat initial condition, since
the Bessel function $I_0(y)$ goes to 1 as $y$ goes to zero. 
We disregard the sub-leading corrections to
the DLA gluon proportional to $\as ln Q^2$ proposed in Ref.\cite{bafo}. 
Integrating expression (\ref{GMF}) over $b_{\perp}$
the Born term then reads
\bea
F_2 = F_2(x, Q^2_0) +  \frac{2}{9 \pi} \int_{Q^2_0}^{Q^2} 
\frac{d Q'^2}{Q'^2} \as(Q'^2) xG^{DLA} (x, Q'^2) \, ,
\label{f2b}
\eea
where we have taken $N_f=3$. The expression (\ref{f2b}) gives the sea
component of $F_2$ generated by the gluon evolution from the initial
virtuality $Q^2_0$ to $Q^2$. Going from expression (\ref{GMF}) to
(\ref{f2b})
we have taken $Q'^2=1/r^2_{\perp}$. Thus, the lower limit $Q_0^2$ 
works as a cut off for the large distances effects over $F_2$. 
These effects are included in $F_2(x, Q^2_0)$,
the value of the
structure function for the virtuality $Q_0^2$. It has a 
nonperturbative origin and takes into account the amount of $q \bar q$
pairs not generated by the perturbative transition 
$g \rightarrow q \bar q$. We will parameterize the initial
structure function by the expression
\bea
F_2(x, Q^2_0)= C_0 x^{-0.08} (1-x)^{10}\, ,
\label{f20}
\eea
where $C_0$ is a constant that adjusts the nonperturbative contribution.
This expression reproduces the soft pomeron behaviour ($ x^{-0.08}$ as $x
\rightarrow \, \infty$) presented by the  $\gamma^*- Nucleon$ cross
section in the low $Q^2$ region\cite{dola}.  
Since we have used $N_f =3$, we should add the charm component $F_2^c$. 
This component is generated  perturbatively from the $\gamma^*- gluon$ fusion
mechanism with the gluon distribution given by the DLA expression (\ref{xgdla}).
This mechanism is discussed in detail in Ref.\cite{grv95}. 
Finally, we
obtain the following expression for $F_2$
\bea
F_2 (x, Q^2)= F_2(x,Q_0^2) + F_{2, \, DLA}^{Born}
+ F_{2, \, DLA}^{c} \, .
\label{f2f1}
\eea
 To fit the expression (\ref{f2f1}) to the HERA data, we have
taken the $F_2$ points which lies in the region
$1\, GeV^2 \, < \, Q^2 \, < \, 100 \, GeV^2$ and $x \, < \, 10^{-2}$, 
where we expect that our DLA approach to SC is valid.
The H1 and ZEUS results were taken from Refs.\cite{h1} and \cite{ze},
respectively.

\begin{figure}[htbp]
\center{\epsfig{file=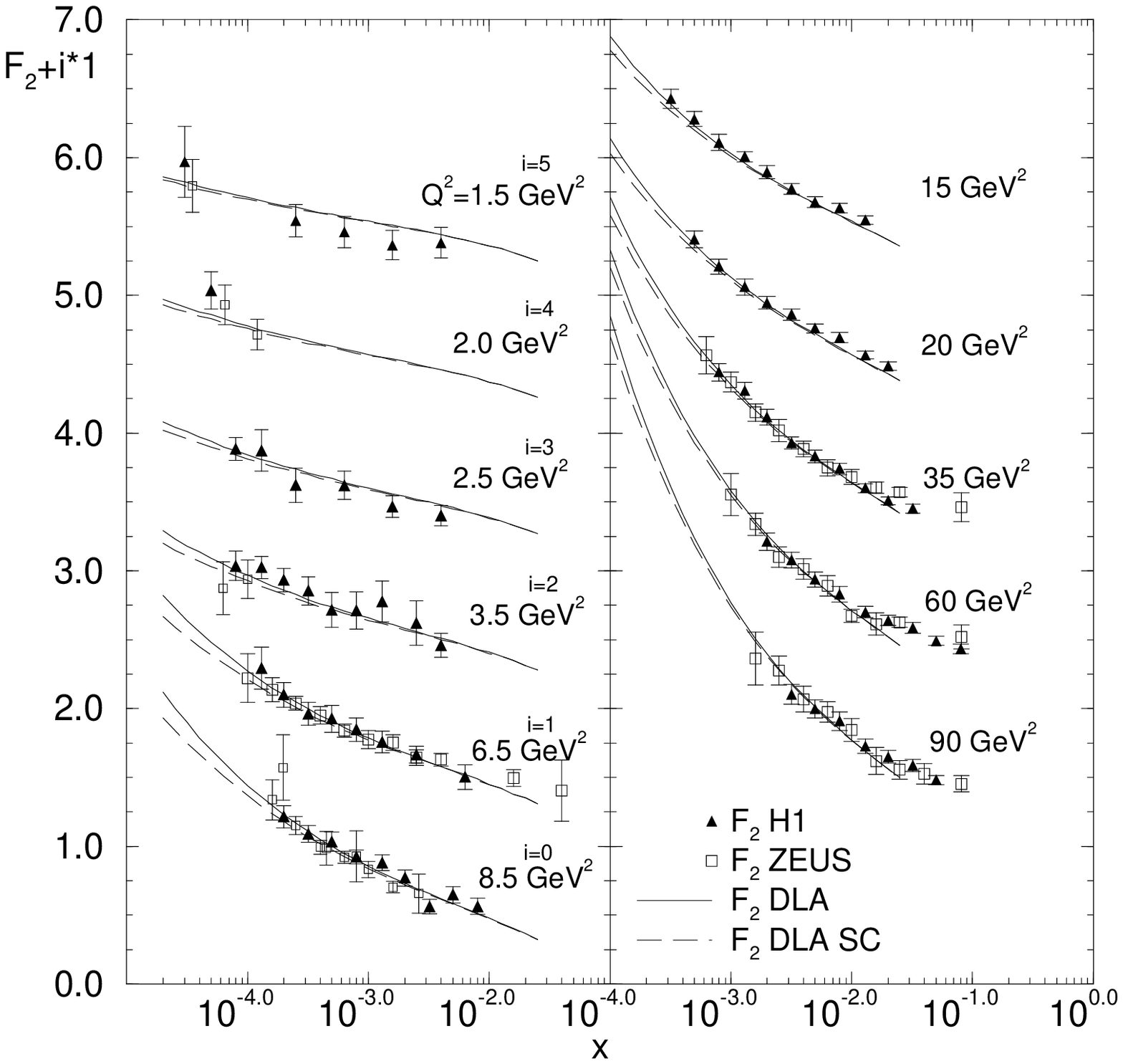,width=120mm} }
\caption{\label{f2dla1} The structure function $F_2$ evolution 
from the scaling violation
mechanism in DLA aproximation as a function of $x$. The solid line 
represents the Born term and the dashed line includes the shadowing
corrections (SC).}
\end{figure}

\begin{figure}[htbp]
\begin{tabular}{c c}
\epsfig{file=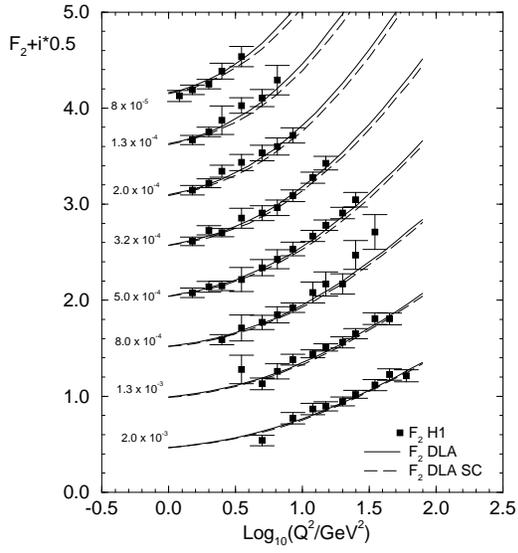,width=90mm} & \epsfig{file=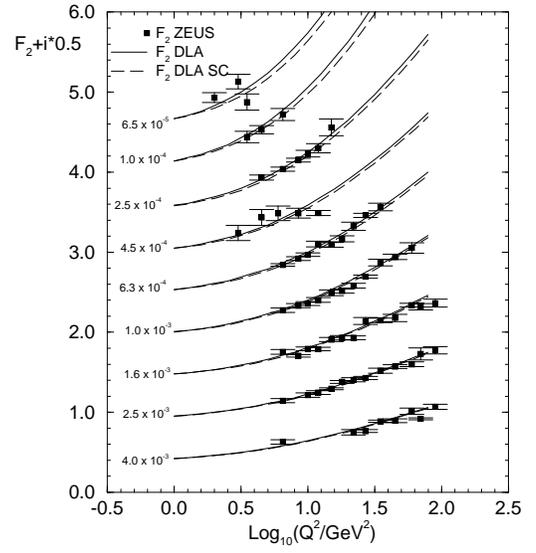,width=90mm} \\
\end{tabular}
\caption{\label{f2dla2} $F_2$ evolution from the scaling
violation mechanism in DLA as a function of  $Q^2$
( for the scaling violation figures, the value of $i$ goes
from $0$ for $x=2.0 \, \times \, 10^{-3}$ to $7$ for
$x=8.0 \, \times \, 10^{-5}$).}
\end{figure}

In figure (\ref{f2dla1}) we present the fit for a subset of
the experimental data.
The parameters used are 
$G_0 = 0.136$, $C_0=0.273$ and  $Q_0^2=0.330 \, GeV^2$.
The values of the parameters were chosen in such a way to
minimize the $\chi^2$, which corresponds to $\chi^2 / d.o.f. = 124/222$. 
We can see from the figure that 
the steep behaviour of the
deep inelastic struture function
is well described by the DLA evolution of the gluon distribution,
regarded we have included enough nonperturbative $q \bar q$ pairs.
With this set of parameters also the $Q^2$ scaling violation
 of $F_2$
can be described, as shown in figure (\ref{f2dla2}). Taking a
small value for the initial virtuality we can generate the
DLA behavior for $Q^2 \approx 2 \, GeV^2$. 
A similar result was obtained
in Ref.\cite{ina}, but in a not completely DLA limit\footnote{The authors
have taken $P_{qg}= z^2 + (1-z)^2$ and the sub-leading factor 
$(t/t_0)^{- \delta}$ with $\delta= (11 +\frac{2 N_f}{27})/\beta_0$}. 
It is important to note that our
aim in this letter is to describe HERA data in a completely 
consistent DLA limit, and
not to provide an overall  fit to existing high energy data. 

Since we have described the data with the Born term of the DLA 
expression (\ref{GMF}),
we can  investigate the amount of shadowing corrections predicted
for $F_2$ on Glauber - Mueller formula of \eq{GMF}.
For that, we substitute the Born term in expression (\ref{f2f1}) by
the full series, which is  taken into account in  
expression (\ref{GMF}). In figure (\ref{f2dla1}) we present the
results for $F_2$ as a function $x$, and in figure (\ref{f2dla2}),
as a function of $Q^2$. As we can see, the shadowing corrections
are important only for very small values of $x$ and moderate values
of $Q^2$. We would like to recall that in \eq{GMF} we put the upper limit
of integration equal to $\frac{1}{Q^2_0}$, or in other words we consider
only the SC which are originated from sufficiently short distance, namely, 
$r_{\perp}\,\leq\, 1/Q_0 \,\approx 0.35 fm$.  In fact, we do not take into 
account the SC at large distances considering that they have been included
in the initial parton distribution of \eq{f20}.

Therefore, we are calculating only
perturbative shadowing. In the kinematic region of present data,
the corrections lie within  the experimental error.

\begin{figure}[hbtp]
\center{\epsfig{file=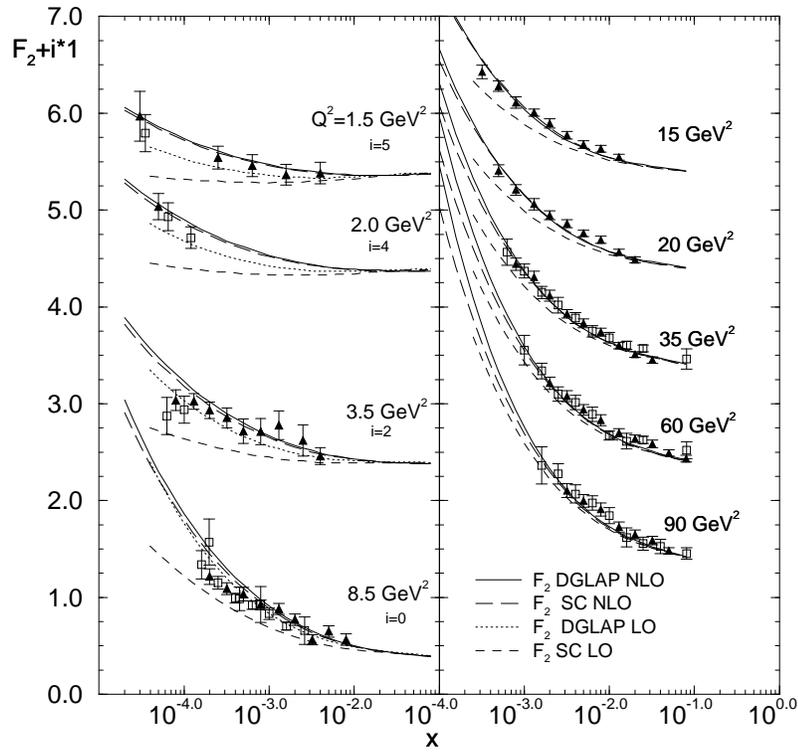,width=120mm} }
\caption{\label{f2dla3} The structure function $F_2$ evolution in LO and NLO 
 as a function of $x$. The Born term (DGLAP evolution) for $F_2$
NLO and LO numerically coincide for $Q^2 \, > \, 10 \, GeV^2$.}
\end{figure}

\begin{figure}[htbp]
\begin{tabular}{c c}
\epsfig{file=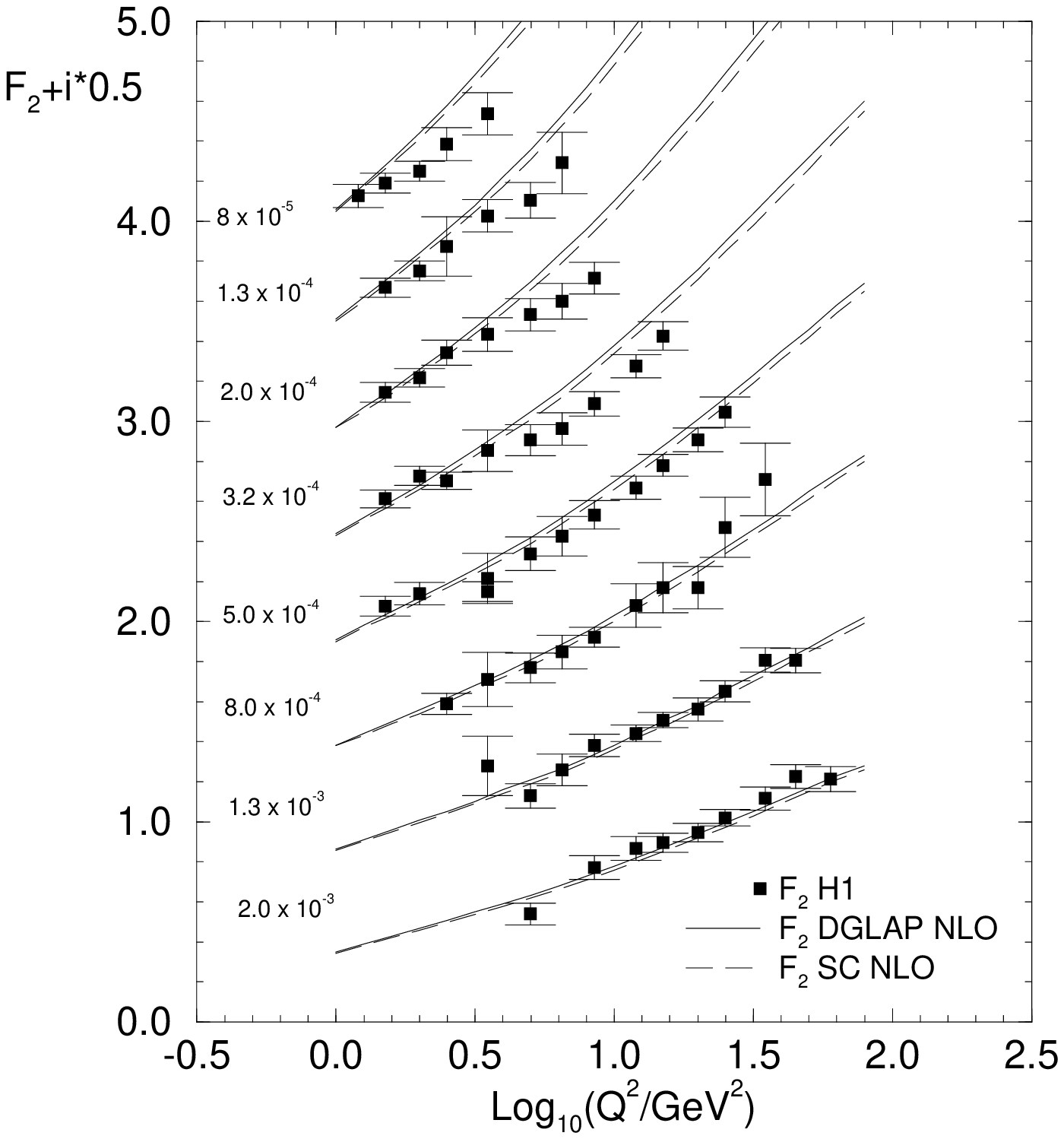,width=90mm} &
\epsfig{file=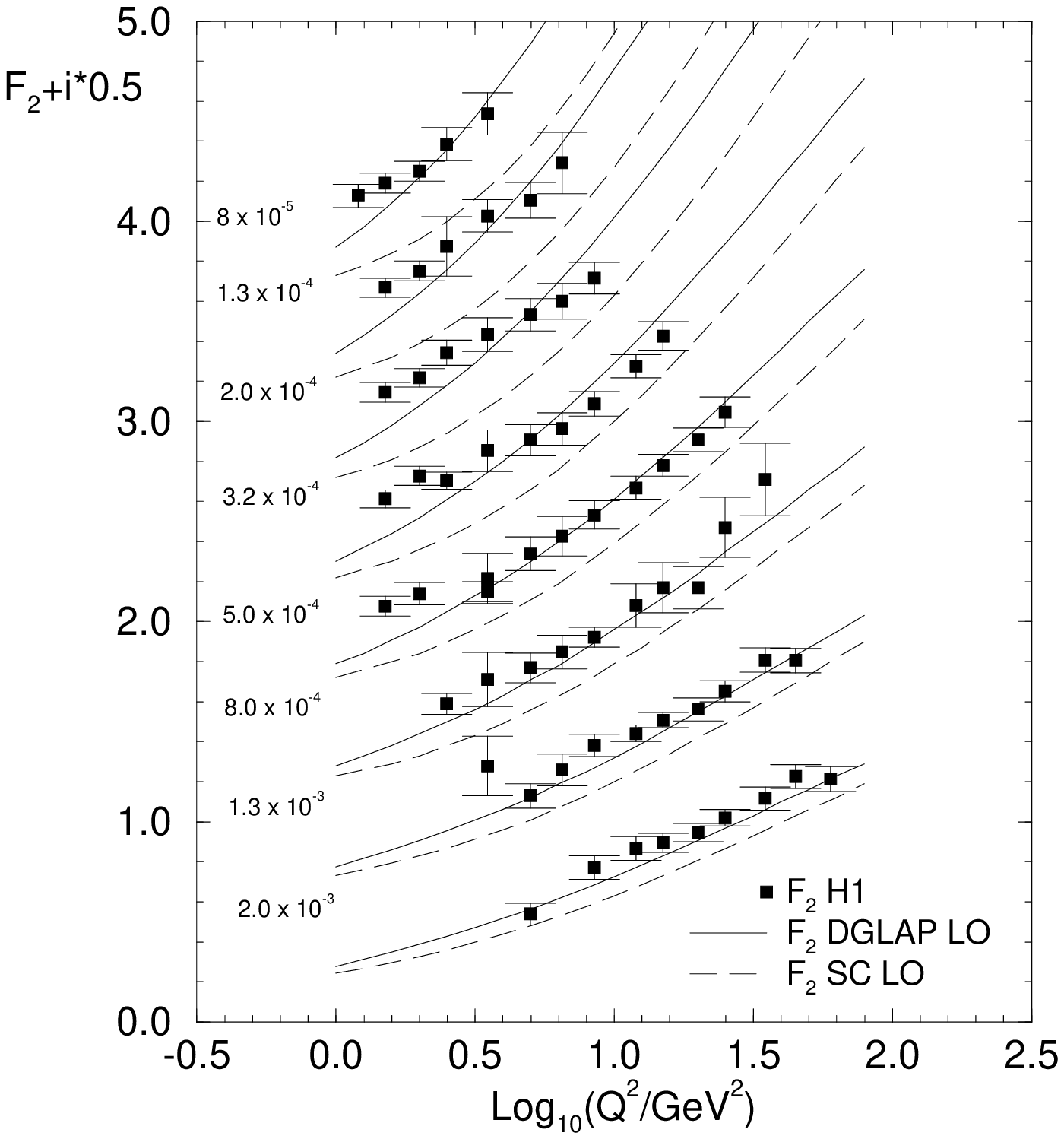,width=90mm} \\
\end{tabular}
\caption{\label{f2dla4} $F_2$ evolution from the scaling
violation mechanism in LO and NLO as a function of  $Q^2$.}
\end{figure}

We plot also in figures (\ref{f2dla3}) and (\ref{f2dla4}) the SC
for $F_2$ predicted by the Glauber approach taking into account
the leading order (LO) and next to the leading order (NLO) gluon.
In both cases, we have used the modified Mueller formula discussed
in Ref.\cite{AGLA}. In this formula, the Born term is taken in 
leading $\as ln Q^2$ aproximation (LLA($Q^2$)), while the correction
term is taken in DLA. For practical purpose, we use the structure
function $F_2$, solution of DGLAP equations,  as the Born term
in expression (\ref{f2f1}). 
We have taken only the  GRV distribution since
those distributions evolve from small virtualities and can be compared
with our DLA approach. We see from the figures that the LO gluon
predicts much more SC to $F_2$. It means that the scaling violation
suffers a stronger modification for the LO gluon when compared to the
simple DLA gluon and to the NLO gluon.

\begin{figure}[hbtp]
\begin{tabular}{c c}
\epsfig{file=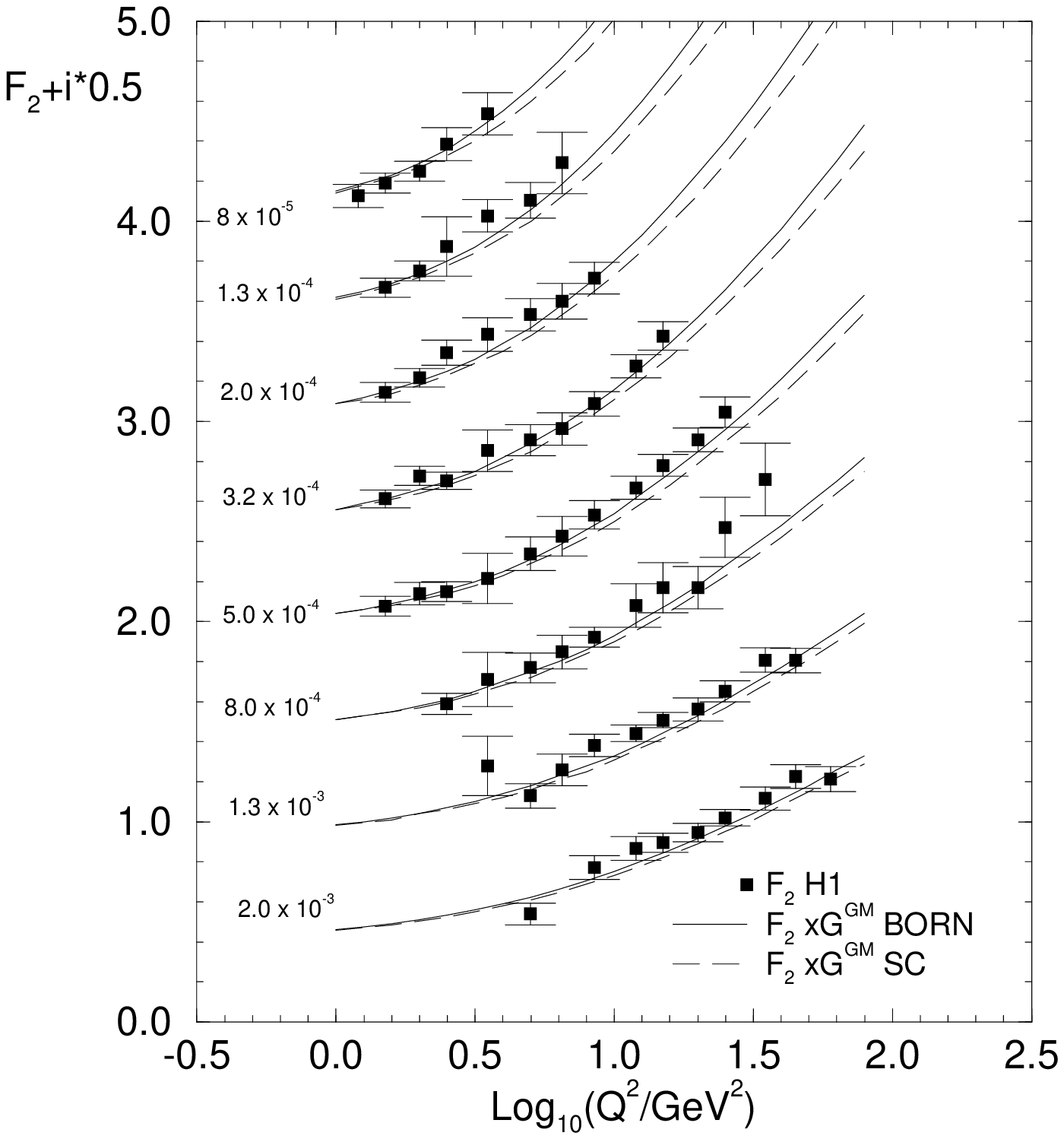,width=90mm} &
\epsfig{file=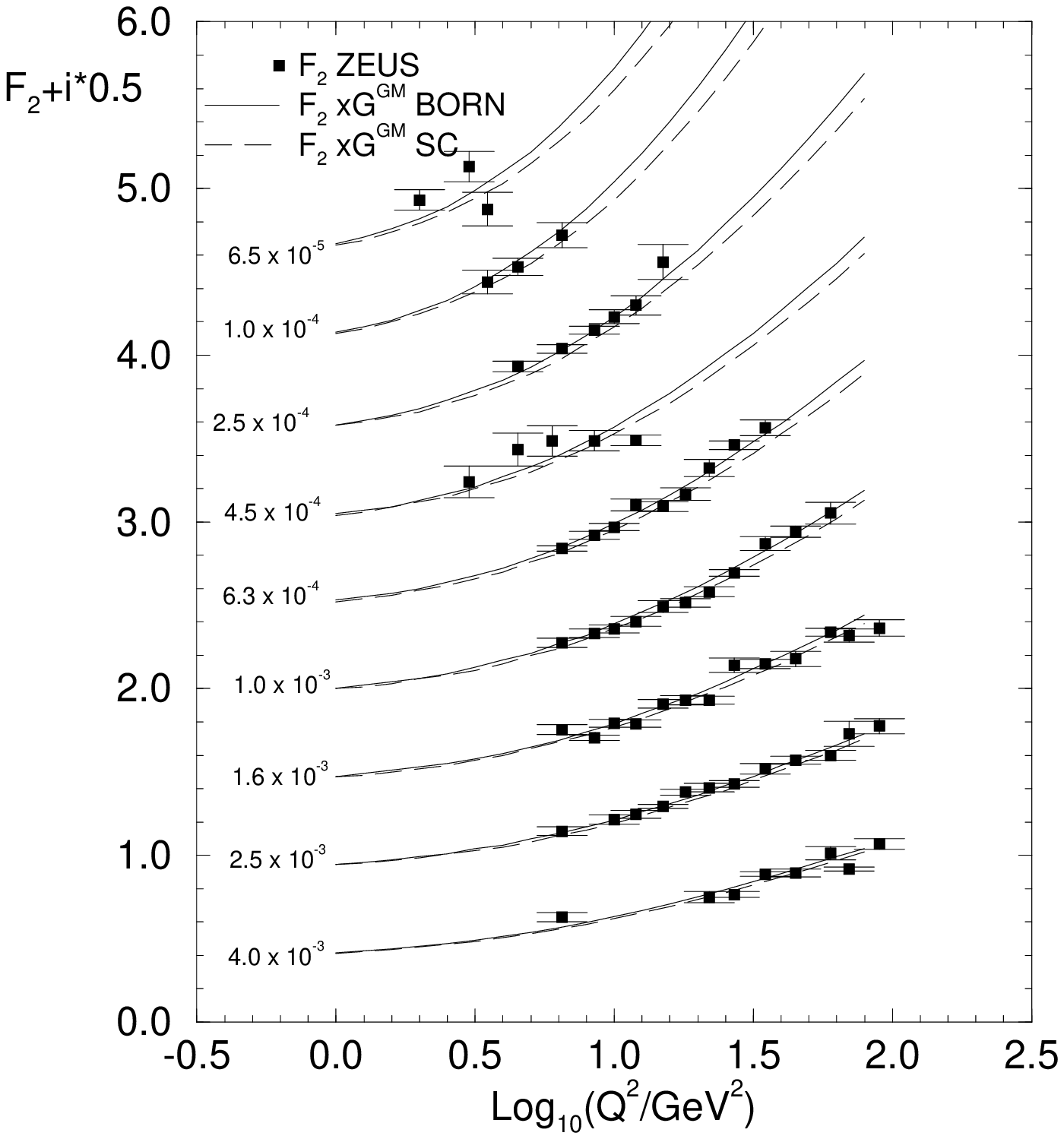,width=90mm}
\\
\end{tabular}
\caption{\label{f2dla6} $F_2$ evolution from the scaling
violation mechanism in DLA with the gluon distribution \protect$xG^{GM}$
given by the Glauber - Mueller approach.}
\end{figure}

The Glauber - Mueller approach cannot be considered as a full description
of the SC, because it was assumed that only quark - antiquark pair
embodies a multi rescatterings with the target. As was shown in Refs.
\cite{AGLA} \cite{AGLP}\cite{GLR} the gluon rescatterings turn out to be
more essential. To demonstrate this fact we calculate here  the Born term
of \eq{GMF} but given by expression:
\bea
F_2 = F_2(x, Q^2_0) +  \frac{2}{9 \pi} \int_{Q^2_0}^{Q^2}
\frac{d Q'^2}{Q'^2} \as(Q'^2) xG^{GM} (x, Q'^2) \, \,
\label{GGM}
\eea
where $xG^{GM}$ is the gluon structure function calculated in the 
Glauber - Mueller approach, namely
\bea
xG^{GM}(x,Q^2)\,\,= \,\,
\frac{2}{\pi^2}\int^{\frac{1}{Q^2_0}}_{\frac{1}{Q^2}}
\,\,\frac{d r^2_{\perp}}{ r^4_{\perp}}
\,\,\int^1_x\, \,\frac{d x'}{x'}\,\,\int^{\infty}_0\,\,d b^2_{\perp}
\,\,\{\,\,1\,\,-\,\,e^{ -
\frac{\Omega_{G}(x',r^2_{\perp},b_{\perp})}{2}}\,\,\}\,\, 
\label{GGMF}
\eea
where the opacity $\Omega_G\,\,=\,\,\frac{9}{4} \Omega$. Expression
(\ref{GGMF}) is the Mueller formula which was discussed
in detail in Ref.\cite{AGLP}. When \eq{GGMF} is included in expression
(\ref{GGM}), the Born term reproduces \eq{f2b}, since the Born term is 
the DGLAP equation in the DLA limit. 
The other terms
take into account the shadowing corrections to the gluon distribution. 
The results are shown in figure (\ref{f2dla6}). Comparing figures 
(\ref{f2dla2}) and (\ref{f2dla6}), one can see that the SC due
to gluon rescattering is bigger then the corrections due to
quark rescattering.

\begin{figure}[hbtp]
\begin{tabular}{c c}
\epsfig{file=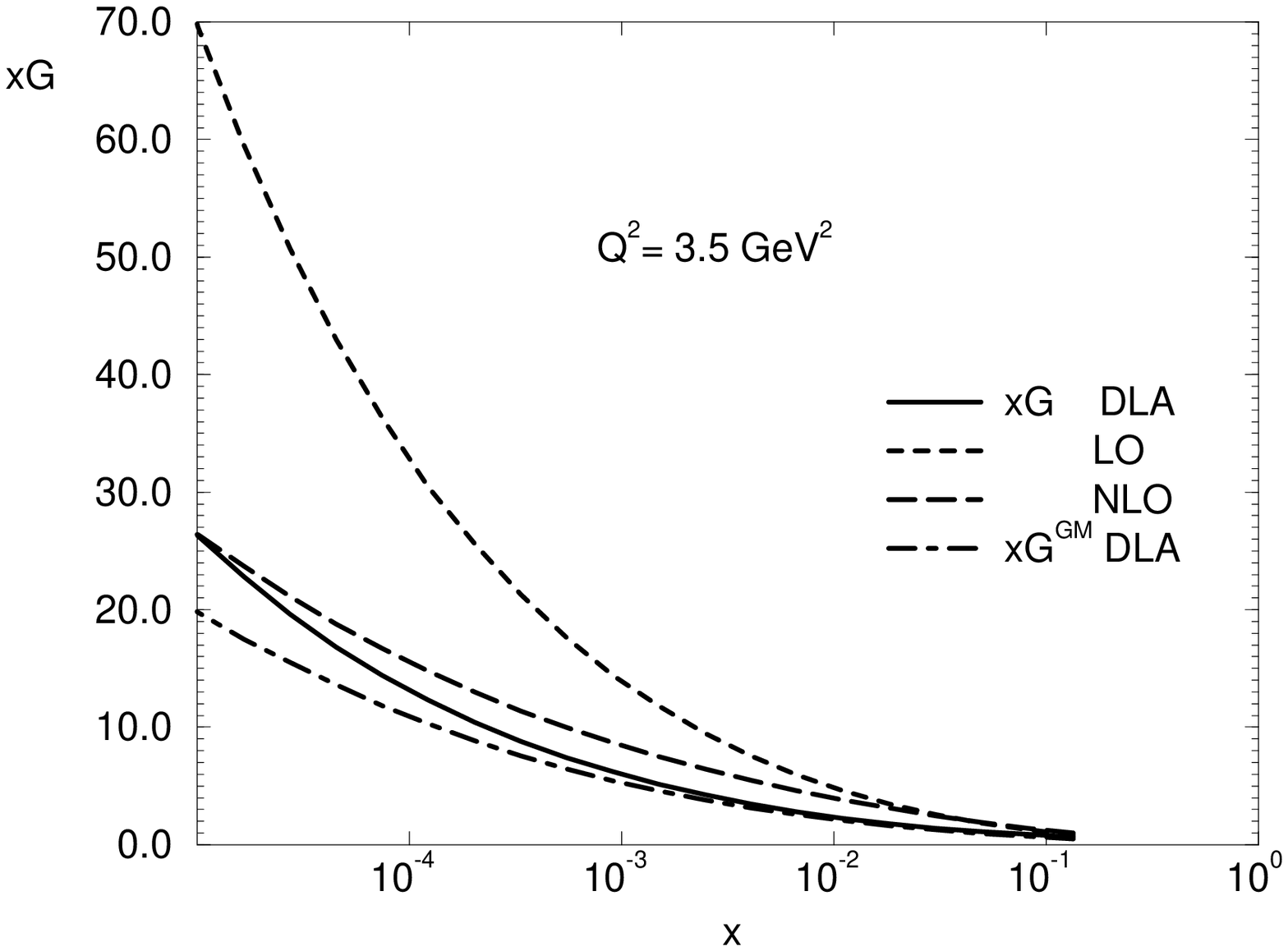,width=80mm} & \epsfig{file=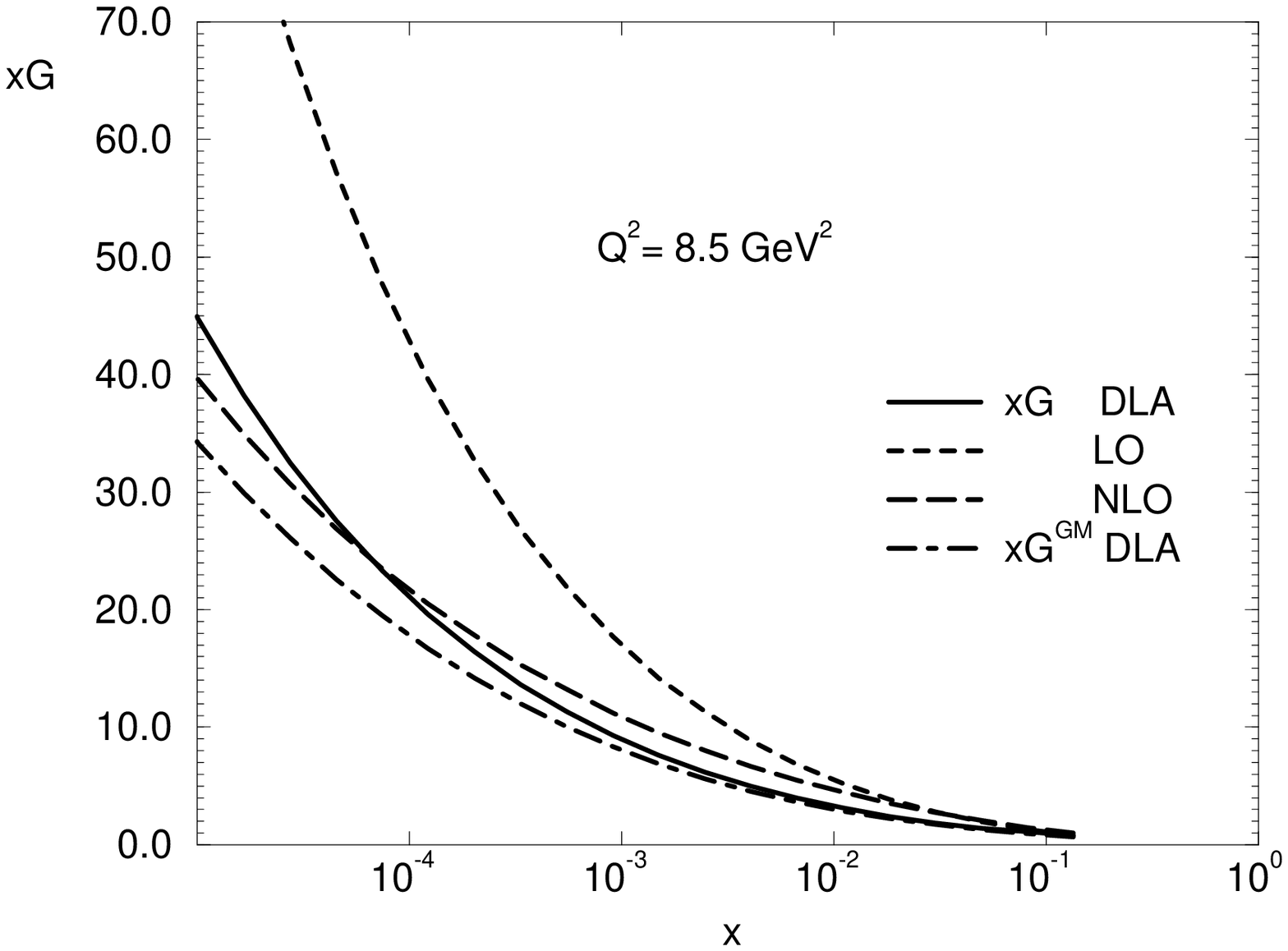,width=80mm}
\\
\end{tabular}
\caption{\label{f2dla5} The gluon distributions in DLA with and without
SC - calculated from the Glauber - Mueller approach, compared
with LO and NLO gluon distribution from GRV95 set.}
\end{figure}

In order to complete our discussion,  we plot in figure (\ref{f2dla5})
the DLA gluon distribution given
by expression (\ref{xgdla}) and the corrected gluon
distribution given by the modified Mueller formula (\ref{GGMF}).
The  LO and NLO gluon
distribution given by the parameterization GRV95 are ploted also.
As we can see, the DLA distribution predicts an amount of gluons
closer to  the NLO DGLAP evolution. It is not a coincidence, since
the NLO GRV distribution has a flat behaviour for $Q^2=0.4 \, GeV^2$
while the LO gluon distribution has already a steep behaviour
in the small $x$ region for this low value of $Q^2$. 

Fig. (\ref{f2dla5}) shows the main conclusion of this letter: the
SC turns out to be big ( about 40\% - 50\% ) in the gluon structure
function but their manifestation in $F_2$  is rather small as we have
discussed ( see  Fig. (\ref{f2dla6}) ).  Comparing also  
figures (\ref{f2dla2}), (\ref{f2dla4}) and (\ref{f2dla6}) we can see
that the SC for $F_2$ have a strong dependence on the ammount of
gluons taken into account in the QCD evolution. This conclusion calls for  
new measurements in the high energy kinematic region more sensitive to the
value of the gluon structure function than the measurements of $F_2$.

ALAF$^{\underline{o}}$ acknowledges CAPES and MBGD acknowledges CNPq for
partial financing. EML thanks E. Gotsman and U. Maor for everyday
discussions on the subject.

~

~
~
\newpage

~

\newpage

\end{document}

%% file: defi.tex
 1
\def\as{\alpha_{\rm S}}

\def\citenum#1{{\def\@cite##1##2{##1}\cite{#1}}}
\def\citea#1{\@cite{#1}{}}

\def\as{\alpha_{\rm S}}

\def\({\left(}
\def\){\right)}

\def\citenum#1{{\def\@cite##1##2{##1}\cite{#1}}}
\def\citea#1{\@cite{#1}{}}

\def\l1vt{\vec{l_{1\perp}}}

\def\rt{r_{\perp}}
\def\bt{b_{\perp}}
\def\rt2{$r^2_{\perp}$}
\def\bt2{$b^2_t$}

\def\jol1{$J_0(\,l_{1\perp}\,r_{\perp}\,)$}

\def\citea#1{\@cite{#1}{}}

\def\beq{\begin{equation}}
\def\eeq{\end{equation}}
\def\bea{\begin{eqnarray}}
\def\eea{\end{eqnarray}}

\def\eq#1{{Eq.~(\ref{#1})}}

\def\bbbz{{\mathchoice {\hbox{$\sf\textstyle Z\kern-0.4em Z$}}
{\hbox{$\sf\textstyle Z\kern-0.4em Z$}}
{\hbox{$\sf\scriptstyle Z\kern-0.3em Z$}}
{\hbox{$\sf\scriptscriptstyle Z\kern-0.2em Z$}}}}
\def\npb#1#2#3{    {\it Nucl. Phys. }{\bf B#1} (19#2) #3}
\def\plb#1#2#3{    {\it Phys. Lett. }{\bf B#1} (19#2) #3}
\def\prd#1#2#3{    {\it Phys. Rev. }{\bf D#1} (19#2) #3}

\def\prl#1#2#3{    {\it Phys. Rev. Lett. }{\bf #1} (19#2) #3}

\def\zpc#1#2#3{    {\it Z. Phys. }{\bf C#1} (19#2) #3}

\def\l{\lambda}